# Addressing User Requirements in Open Source Software: The Role of Online Forums

**Arif Raza*\***

Department of Computer Software Engineering, National University of Sciences and Technology, Islamabad, Pakistan
**arif_raza@mcs.edu.pk**

**Luiz Fernando Capretz**

Department of Electrical & Computer Engineering, University of Western Ontario, London, Canada
**lcapretz@uwo.ca**

## Abstract

User satisfaction has always been important in the success of software, regardless of whether it is closed and proprietary or open source software (OSS). OSS users are geographically distributed and include technical as well as novice users. However, it is generally believed that if OSS was more usable, its popularity would increase tremendously. Hence, users and their requirements need to be addressed in the priorities of an OSS environment. Online public forums are a major medium of communication for the OSS community. The research model of this work studies the relationship between user requirements in open source software and online public forums. To conduct this research, we used a dataset consisting of 100 open source software projects in different categories. The results show that online forums play a significant role in identifying user requirements and addressing their requests in open source software.

**Category:** Human computing

## I. INTRODUCTION

Throughout the design, implementation, and deployment of software, different quality attributes, such as availability, variability, performance, maintenance, testability, modifiability, scalability and usability, are considered with some having more preference over others, depending upon the nature of the application. Usability is one of those quality attributes of a software architecture design, which is always there to be considered, no matter what type of system is going to be designed. This defines the scope of this study. We are focusing on user require-ments and the usability of open source software (OSS). The International Organization for Standardization and the International Electro Technical Commission ISO/IEC 9126-1 [1] places software quality attributes into six categories: functionality, reliability, usability, efficiency, maintainability, and portability. In the standard, usability is defined as "*the capability of the software product to be understood, learned, used and attractive to the user, when used under specified conditions.*"

The use of free and open source software has increased significantly in recent years, mainly due to the accessibility and availability of the Internet. However, software

---

developers and programmers are not the sole users of OSS anymore; the number of non-technical and novice computer users is growing at a fast pace, highlighting the necessity of understanding and addressing their requirements and expectations [2]. Expectations and requirements of these users might not be the same as those of experts and need to be addressed in further studies. OSS is no longer a limited arena for OSS developers and technical contributors alone.

Raza et al. [3], in their empirical study, maintained that the requirements and expectations of OSS users are not the same as before, when software developers were the only OSS users. According to them, in order to improve OSS product quality and usability, OSS designers and developers should comprehend that they are not the eventual users of their project. Bodker et al. [4] also suggest that the right user requirements can help avoiding "*developer-centric*" OSS systems and improve performance in the context of usability standards. Cetin and Gokturk [5] acknowledge that little attention has been paid to users' requirements, even though it is acknowledged as a significant factor. Considering usability as a non-functional requirement, the authors opine that it can only be quantified indirectly through techniques, such as heuristic evaluations, prototyping and cognitive evaluation. They further believe that poor usability testing is mainly due to the unavailability of proper usability requirements.

Online forums are the main communication platforms for the users, developers and other contributors of the OSS community. In this study, the role of online forums is empirically examined in relation to the identification and management of user requirements. A dataset of 100 OSS projects of different categories is used to study the research model of this empirical study.

## II. LITERATURE REVIEW

According to Hepting et al. [6], "an *e-mail archive for the user groups; requested features in the bugzilla; and the feature poll on the project wiki*" are considered as main sources of user requirement collection for open source projects. Molina and Toval [7] observed that the recent failure of many web engineering projects is mainly due to inadequate collection of requirements and metrics of certain quality attributes. They believe that usability needs to be given special importance in the early stages of development. The authors proposed a requirements meta-model which defines the typical elements that participate in Web information system requirements elicitation. Subramaniam et al. [8] also observed a positive association between a restrictive OSS license and the interests of non-developer users. They concluded that the success measures such as "*developer interest, user interest and project activity*" are inter-related. Moreover, they believe that users are more likely to download software in its later stage of development due to its improved usability.

Schwartz and Gunn [9] maintained that for the OSS community, "*usability still appears to be an afterthought.*" They considered complex interfaces of OSS as hindrances to their use by non-technical end users. They advocated the necessity of usability awareness among OSS developers. De Groot et al. [10] observed that not only has the participation grown in OSS development, but number of users and their requirements have increased tremendously as well in the last decade. The authors mentioned that although dozens of OSS applications are available, which differ in quality features and requirements, the end user finds it very difficult to choose one for a given problem. Juristo [11] also upholds a relationship between software design and usability. She thus recommended considering usability features in the requirements stage. However, it was also observed that the discovery and documentation of such features are not trivial for developers.

Lee et al. [12] carried out an empirical study to measure OSS success. In their proposed model, they identified five determinants for OSS success: software quality, community service quality, OSS use, user satisfaction and individual net benefits. The authors maintained that "*the users need support, cooperation, and assistance during both the development and post-development phases.*" The results of the study indicate that the software quality and user satisfaction are two factors that play major roles in determining OSS usage. Gallego et al. [13] also observed that little attention had been paid to the issue of user acceptance as far as OSS is concerned. They identified and studied the relationship of different factors having an effect on users' attitudes towards OSS adoption. Based on the outcome of the study, the authors emphasized that OSS developers need to realize that OSS adoption is dependent on users' acceptance. They further concluded that useful and easy-to-use OSS is a stimulating factor for significant use of OSS. Furthermore, "*organizations and users should consider criteria such as software quality, systems capability and software flexibility for selecting the most adequate OSS.*"

Stam [14] examined the implication of user activities over innovative performance of the organizations using OSS. It was observed that organizations' contribution to OSS community development is dependent upon end user requirements in addition to other factors, such as source code and technical knowledge. Scacchi [15] comparatively studied the requirements engineering process to characterize how the requirements for OSS are carried out in different OSS projects of different communities. It was found that OSS development involves a social and technical relationship which is not there in the case of traditional software development. The study advocated that "*open software requirements artefacts might be assessed* in terms of virtues such as encouragement of community building, freedom of expression and multiplicity of expression with software informalisms, readability and ease of navigation, and implicit versus explicit structures for organising, storing and sharing open software requirements." The open source approach is thus bringing a new way of requirements gathering, construction, deployment, and evolution of software.

Breu et al. [16] focused on collaboration between the developers and the users of OSS through bug tracking systems. While studying 600 bug reports, the authors found out that in OSS, the role of users is not limited to simply bug reporting. They believe that *an integration and active participation of users in bug tracking will result in bugs being fixed faster and more efficiently.* They maintained that although user involvement is vital for effective bug reporting, there is a need to have efficient tools to facilitate users [17].

Based on the above discussion, it is evident that online forums have a crucial role in the development and maintenance of OSS. Online forums provide a major platform as far as users' involvement is concerned. In particular, we intend to explore whether users' requirements are being addressed by OSS developers. It is also investigated whether public online forums assist in identifying and addressing users' requirements.

## III. RESEARCH MODEL AND HYPOTHESES

In this study, we present a research model for analyzing the relationship between user requirements and the online forums of OSS projects. The theoretical model that will undergo empirical testing is presented in Fig. 1. Although it is generally believed that OSS developers refine their work through other contributors' input, in this study, we are particularly focusing on whether Support Request Forums and Feature Request Forums assist in identifying and addressing end users' demands. The term "Support Requests Forum" refers to a forum that deals with any activity, such as installation help, documentation, information retrieval, usability, etc., whereas the phrase "Feature Requests Forum" is used for a platform that deals with changes in requirements to add more requested functionality [18].

Our aim is to investigate the answer to the following research question (RQ):

*RQ: Do online forums assist in addressing user requirements in OSS projects?*

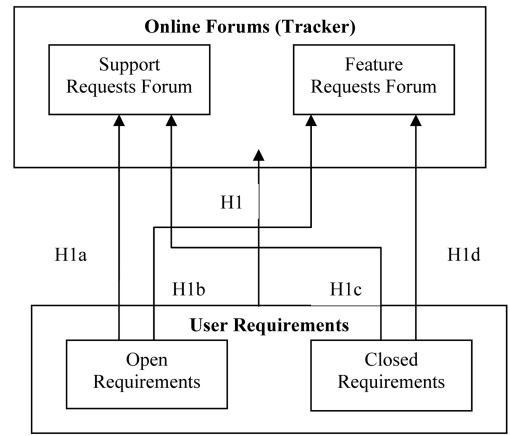

**Fig. 1.** Research model.

The term "Open Requirements" is used as a user request or requirement that has been reported but has not yet been addressed, whereas "Closed Requirements" refers to a user requirement that has been reported and taken care of. In order to empirically investigate the research question, five hypotheses are presented in Table 1.

## IV. RESEARCH METHODOLOGY

Data was collected from 100 projects on sourceforge.net, a popular open source software project repository. The dataset covers the ten most downloaded projects from each of the categories of communication, software development, Internet, games & entertainment, text editors, scientific/engineering, database, education, multimedia, and formats & protocols. The data was collected from statistics of tracker traffic of each project for the time period of one year (September 2010–August 2011). The first filtration activity removes the data of all projects which have either total requirements of zero or no online

**Table 1.** User requirements hypotheses

| Hypothesis # | Statement |
| --- | --- |
| H1 | Public online forums help in identifying and addressing user requirements in OSS projects. |
| H1a | The open requirements are positively related with the number of "Support Requests" in online forums. |
| H1b | The open requirements are positively related with the number of "Feature Requests" in online forums. |
| H1c | The closed requirements are positively related with the number of "Support Requests" in online forums. |
| H1d | The closed requirements are positively related with the number of "Feature Requests" in online forums. |

OSS: open source software.

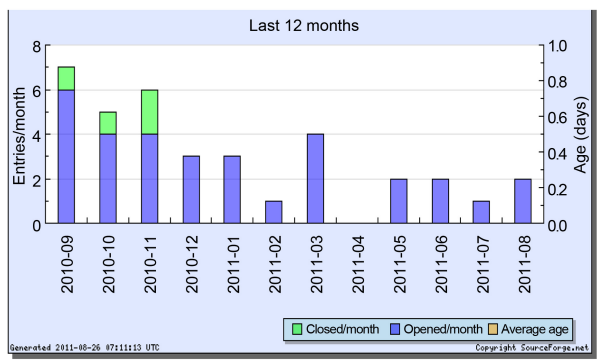

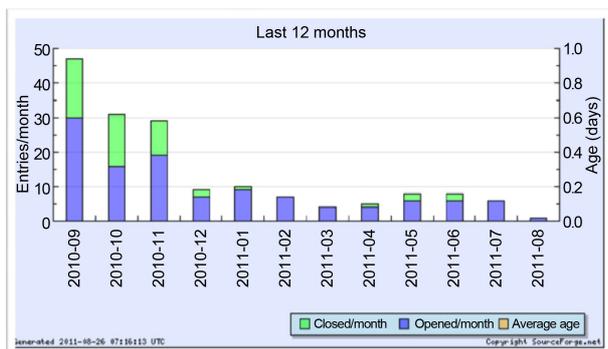

**Fig. 2.** Sample tracker statistics of open source software projects.

forums. This reduces the dataset to 72 projects.

The maximum total open requirements/user requests (7,070) were found in the category of scientific/engineering but none of them were closed. The maximum requests (54.10%) addressed were observed in a text editor project. The maximum open support requests (2,804) were also found in the category of scientific/engineering, whereas a project in the Internet category had the highest percentage (63.64%) of the closed support requests. The category of "games/entertainment" had the maximum number of open feature requests (163) in one project. The maximum closed feature requests (51.85%) were observed in an Internet project. Fig. 2 presents the one-year tracker traffic of open and close requests in online forums of two sample projects.

## A. Data Analysis Procedure

We conducted tests for hypothesis H1, and its sub-hypotheses H1a, H1b, H1c, and H1d. Initially, we divided the data analysis activity into three phases. In Phase I, we conducted tests for the hypotheses using parametric statistics by computing the Pearson correlation coefficients, and in Phase II, we utilized non-parametric statistics through the Spearman correlation coefficients. Due to the relatively small sample size, both parametric as well as non-parametric statistical approaches were used to reduce the threat to external validity. The results are presented in Table 2.

The Pearson correlation coefficient and t-test were examined between variables involved in the hypotheses H1a, H1b, H1c, and H1d. The Pearson correlation coefficient between open requirements and support requests in the public forums was positive (0.985) at $p = 0.000$, and thus provided a justification to accept the hypothesis H1a. The hypothesis H1b was also accepted based on the Pearson correlation coefficient (0.591) at $p = 0.000$, between open requirements and number of feature requests in the online forums. The correlation coefficient of 0.019 at $p = 0.846$ was observed between the closed requirements and number of support requests in the online forums, and hence H1c was rejected. The hypothesis H1d was accepted based on the Pearson correlation coefficient (0.646) at $p = 0.000$, between closed requirements and feature requests in the online forums. Hence, it was observed and that hypotheses H1a, H1b, and H1d were found to be statistically significant and were accepted. However, the hypothesis H1c was not supported by our parametric analysis and was hence rejected.

In Phase II, we conducted a non-parametric statistical technique using the Spearman correlation coefficient to test our hypotheses. Hypothesis H1a was statistically significant at $p = 0.000$ with a Spearman correlation coefficient of 0.421. A positive association was observed between open requirements and feature requests (Spearman 0.701 at $p = 0.001$) in the online forums. H1c, which deals with the closed requirements and support requests

**Table 2.** Empirical analysis results

| | Requirement | | | |
|---|---|---|---|---|
| | Pearson correlation | | Spearman correlation | |
| | Open | Closed | Open | Closed |
| Support requests | 0.985 (p = 0.000) | 0.019 (p = 0.849) | 0.421 (p = 0.000) | 0.272 (p = 0.006) |
| Feature requests | 0.591 (p = 0.000) | 0.646 (p = 0.000) | 0.701 (p = 0.000) | 0.688 (p = 0.000) |

Significant at p < 0.005.

**Table 3.** Structural test results

| | Requirement | |
|---|---|---|
| | Open | Closed |
| **Support requests** | | |
| Coefficient | 0.27 | 0.19 |
| $R^2$ | 0.177 | 0.074 |
| F-ratio | 21.15 | 7.83 |
| p-value | 0.000 | 0.006 |
| **Feature requests** | | |
| Coefficient | 0.59 | 0.64 |
| $R^2$ | 0.491 | 0.473 |
| F-ratio | 94.47 | 88.13 |
| p-value | 0.000 | 0.000 |

Significant at $p < 0.005$.

in the online forums, could not be accepted due to the observed Spearman coefficient of 0.272 at $p = 0.006$. The Spearman correlation of 0.688 at $p = 0.000$ was observed for H1d and was thus accepted. Hence, it was observed that hypotheses H1a, H1b, and H1d were statistically significant and accepted. However, like in Phase I, H1c was rejected due to the statistical analysis results.

Finally, in Phase III of hypotheses testing, we used the partial least square (PLS) technique to cross validate the results of Phase I and Phase II. We tested the hypotheses H1a, H1b, H1c, and H1d by examining their direction and significance. The hypothesis involves two variables, so in PLS, we placed one variable (support requests or feature requests) as the response variable and another (open requirements or closed requirements) as the predicate. Table 3 reports the results of the structural tests of the hypotheses. It contains observed values of the path coefficient, $R^2$, and F-ratio.

The results are displayed in Table 3. The PLS technique is especially useful in situations involving complexity, non-normal distribution, low theoretical information, and small sample size [19, 20]. All statistical calculations were performed using Minitab 16 software.

The path coefficient of open requirements (H1a) was found to be 0.27 with $R^2$ (0.177), and the F-ratio (21.15) was significant at $p = 0.000$. Open requirements (H1b) also had a positive path coefficient of 0.59 with $R^2$ (0.491) and an F-ratio of 94.47 at $p = 0.000$ with the number of feature requests. Closed requirements with support requests had a path coefficient of 0.19, $R^2$ (0.074), and F-ratio (7.83) at $p = 0.006$, which did not support hypothesis H1c. Closed requirements and the feature requests had a path coefficient of 0.64, $R^2$ (0.473), F-ratio (88.13) at $p = 0.000$, and had the same direction as proposed in H1d. Overall, the hypotheses H1a, H1b, and H1d showed significance at $p < 0.005$ with positive path coefficients and were in the same direction as proposed, but H1c was not supported in this phase as well.

## V. STUDY DISCUSSION AND CONCLUSION

The open source community is not limited to developers and programmers any more. Users from all over the world are using OSS, with their different cultural backgrounds and unique requirements and expectations. In this research, we studied the relationship between OSS users' requirements for different supports and features and their response through public online forums.

The results of our study demonstrate positive correlation between the open user requirements, support requests, and feature requests in the online forums. Similar correlation has also been observed between the closed user requirements and the number of feature requests. Our analysis, however, did not support the correlation between the closed user requirements and the support requests. This necessitates addressing support requests per requirements and expectations of OSS users.

The overall results of our statistical investigation indicate a positive correlation between user requirements and the role of online forums. The outcome of the study supports hypothesis H1 and our assumption that online public forums help in addressing user requirements in OSS projects. This, in turn, provides an answer to the RQ as well.

As with any empirical studies, this research study has limitations of its own. Construct validity, internal validity, external validity and reliability are four criteria of validity in an empirical study [21]. Threats to the external validity limit the researcher's ability to generalize the experimental results to industrial practice [22]. Furthermore, there have been ethical issues related to empirical studies in software engineering [23, 24].

In this study, we used a random sampling technique to address threats to external validity. The data has been collected from the most downloaded projects. Additionally, we retrieved data from the most active and well-known OSS reporting website, sourceforge.net, which includes a large number of projects. Moreover, we followed recommended ethical principles to avoid violation of any of the recommended experimental ethics.

Addressing users' requirements is one of the major challenging options for open source projects. The objective of this study was to empirically analyze the association between user requirements in OSS and support through online public forums. The empirical results of this study support the hypotheses that public online forums assist in addressing user requirements in OSS projects. The study results, however, suggest that user requirements need to be more thoroughly addressed by using support request forums.

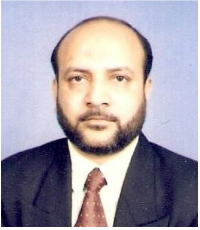

### Arif Raza

Arif Raza received his M.Sc. (1994) in Computing Science from the University of London (UK) and a Ph.D. (2011) in Software Engineering from the University of Western Ontario, Canada. He has several years of teaching experience in Computer Science and Software Engineering. He has authored and co-authored several research articles in peer-reviewed journals and conference proceedings. His current research interests include empirical investigation about open source usability improvement, human computer interaction (HCI), and human factors in Software Engineering.

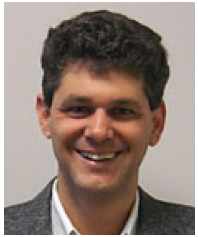

### Luiz Fernando Capretz

Luiz Fernando Capretz has vast experience in the software engineering field as a practitioner, manager and educator. He is currently the Assistant Dean for IT and e-Learning, and former Director of the Software Engineering Program at Western. He has authored over 200 academic papers on software engineering in leading international journals and conference proceedings, and co-authored two books: *Object-Oriented Software: Design an Maintenance* published by World Scientific and *Software Product Lines* published by VDM-Verlag. His current research interests are software engineering, human aspects of software engineering, software product lines, and software engineering education. He is a senior member of IEEE, a distinguished member of the ACM, a MBTI Certified Practitioner, and a Certified Professional Engineer in Canada (P.Eng.).